\documentclass{article}
\usepackage{arxiv}
\usepackage[utf8]{inputenc}
\usepackage[T1]{fontenc}

\usepackage[square,numbers,sort&compress]{natbib}
\usepackage{fontawesome5}
\usepackage{hyperref}
\usepackage{url}
\usepackage{booktabs}
\usepackage{amsfonts}
\usepackage{nicefrac}
\usepackage{microtype}
\usepackage{lipsum}
\usepackage{graphicx}
\usepackage{threeparttable}
\usepackage{doi}
\usepackage{multirow}
\usepackage{multicol}
\usepackage{xcolor}
\usepackage{subcaption}
\usepackage[normalem]{ulem}
\usepackage{amsmath,amsfonts,bm}
\usepackage{amssymb,amsthm}
\usepackage{enumitem}
\usepackage{adjustbox}
\usepackage{blindtext}
\usepackage{float}
\usepackage{array}
\usepackage{makecell}
\usepackage[perpage]{footmisc}
\usepackage{wrapfig}
\usepackage{tcolorbox}
\usepackage{CJKutf8}

\usepackage{tcolorbox}
\tcbuselibrary{breakable, skins} 
\usepackage{CJKutf8}
\usepackage{tikz} 

\definecolor{casebg}{RGB}{236, 241, 245} 
\definecolor{thinkgray}{RGB}{80, 80, 80}

\newtcolorbox{CaseStudyBox}{
  colback=casebg,
  colframe=black,
  boxrule=1.2pt,
  arc=4mm,
  boxsep=8pt,
  left=10pt, right=10pt, top=10pt, bottom=10pt,
  breakable,
  enhanced,
  parbox=false,  
}

\usepackage{cleveref}

\setlist[itemize,1]{leftmargin=\dimexpr 18pt}
\setlist[enumerate,1]{leftmargin=\dimexpr 18pt}

\title{
\raisebox{-0.1\height}{\includegraphics[width=0.04\textwidth]{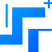}} %
Step-Audio-R1{.}5 Technical Report
}

\author{\vspace{1em} StepFun-Audio Team\\
\vspace{1em}
\faGithub \quad \href{https://github.com/stepfun-ai/Step-Audio-R1}{StepAudio R1.5 Official Github Page}\\
}

\renewcommand{\headeright}{StepFun-Audio Team}
\renewcommand{\undertitle}{}
\renewcommand{\shorttitle}{}

\begin{document}
\large

\maketitle

\begin{abstract}
Recent advancements in large audio language models have extended Chain-of-Thought (CoT) reasoning into the auditory domain, enabling models to tackle increasingly complex acoustic and spoken tasks. To elicit and sustain these extended reasoning chains, the prevailing paradigm—driven by the success of text-based reasoning models—overwhelmingly relies on Reinforcement Learning with Verified Rewards (RLVR). However, as models are strictly optimized to distill rich, continuous auditory contexts into isolated, verifiable text labels, a fundamental question arises: \textbf{are we fostering true audio intelligence, or merely reducing a continuous sensory medium into a discrete puzzle?} We identify this as the \textbf{"verifiable reward trap."} While RLVR yields remarkable scores on standardized objective benchmarks, it systematically degrades the real-world conversational feel of audio models. By prioritizing isolated correctness over acoustic nuance, RLVR reduces dynamic interactions to mechanical "answering machines," severely compromising prosodic naturalness, emotional continuity, and user immersion, particularly in long-turn dialogues. To bridge the gap between mechanical objective verification and genuine sensory empathy, we introduce Step-Audio-R1.5. Marking a paradigm shift toward Reinforcement Learning from Human Feedback (RLHF) in audio reasoning. Comprehensive evaluations demonstrate that Step-Audio-R1.5 not only maintains robust analytical reasoning but profoundly transforms the interactive experience, redefining the boundaries of deeply immersive long-turn spoken dialogue.
\end{abstract}

\section{Introduction}
\label{sec:introduction}
\begin{figure}[t]
    \centering
    \includegraphics[width=\linewidth]{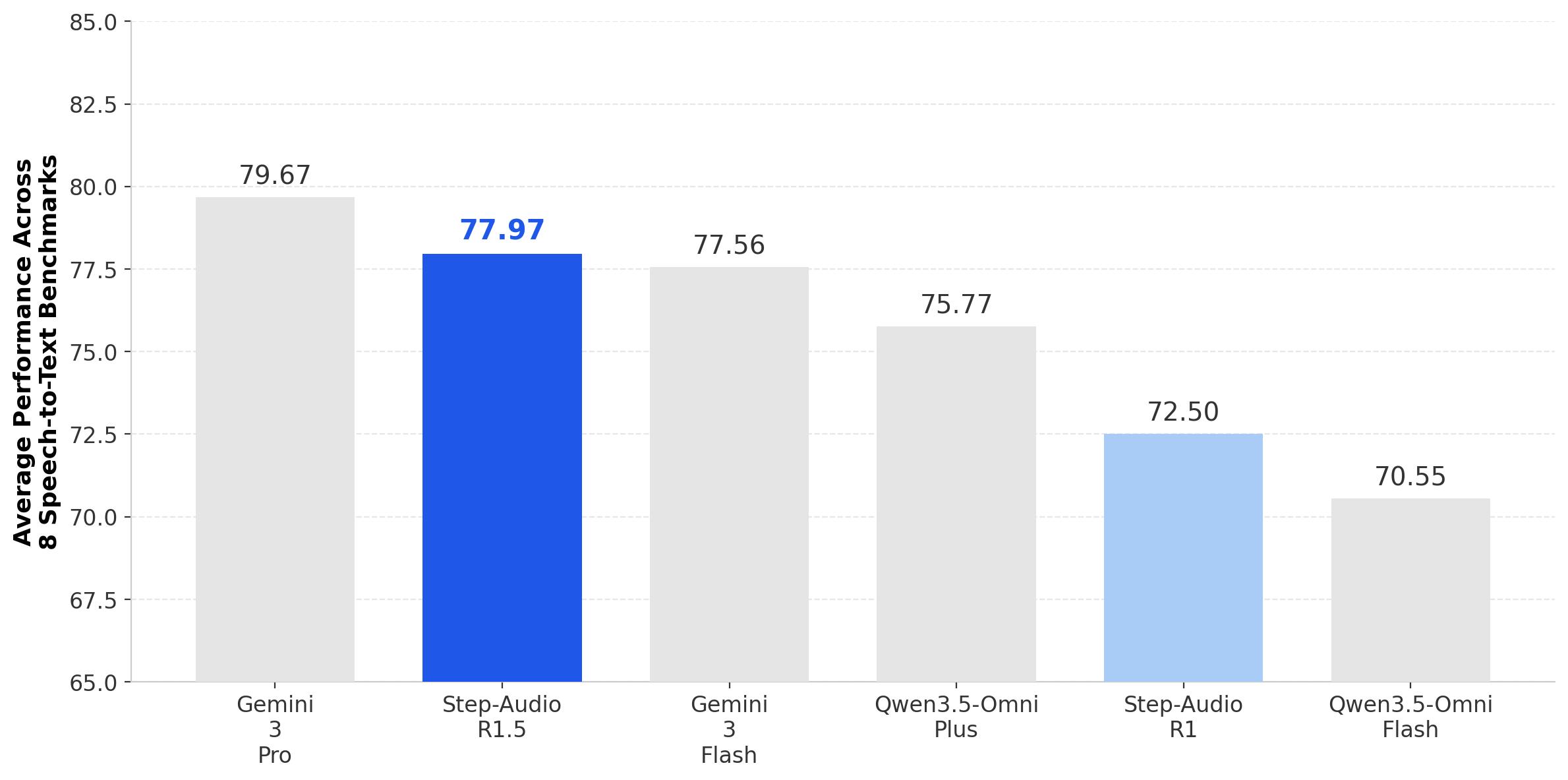}
    \caption{
        \textbf{Aggregate Performance across Speech-to-Text Benchmarks.} 
        The average score represents the holistic capabilities of each model computed over 8 distinct reasoning and perception benchmarks, including Audio MultiChallenge, Big Bench Audio, MMSU, MMAU, Spoken MQA, Step-Caption, Step-DU, and Step-SPQA. 
        Step-Audio-R1.5 substantially outperforms its predecessor and remains highly competitive with state-of-the-art commercial systems such as Gemini 3 Pro.
    }
    \label{fig:teaser_avg_performance}
\end{figure}

Chain-of-Thought reasoning has substantially advanced large language models. By decomposing complex problems into explicit intermediate steps, models such as OpenAI o1~\cite{jaech2024openai} and DeepSeek-R1~\cite{guo2025deepseek} have achieved human-level performance in mathematical olympiads, competitive programming, and scientific inquiry. Central to this progress is Reinforcement Learning with Verified Rewards (RLVR)~\cite{guo2025deepseek}, a training paradigm that reinforces extended reasoning chains using binary, automatically checkable correctness signals, thereby bypassing the need for a learned reward model.

Efforts to transplant this recipe into the auditory domain are accelerating. A growing body of large audio language models~\cite{xu2025qwen3,zhang2025mimo,wu2025step} applies CoT reasoning to speech, music, and environmental sound, with early RLVR-trained variants~\cite{xie2025audio,rouditchenko2025omni,tian2025step} reporting strong results on objective tasks such as speech question answering and acoustic scene tagging. However, these benchmarks share a critical structural limitation: the temporally extended audio input is ultimately reduced to a single discrete label---a category, a number, or a short factual string. Consequently, RLVR can only reward the model for producing that specific label, leaving it structurally blind to prosodic naturalness, emotional continuity, and conversational coherence. We term this the \textbf{verifiable reward trap}: the optimization objective strictly selects for isolated answer accuracy while ignoring the nuanced qualities that determine user experience in real-world deployment.

The empirical consequence of this trap is consistent and reproducible. Under prolonged RLVR training, models become increasingly accurate on held-out test sets yet increasingly unnatural to interact with; responses grow terse, mechanical, and emotionally flat. In multi-turn spoken dialogues, where users expect not merely correct answers but genuine conversational flow, the model often degenerates into a literal ``answering machine''---technically accurate but experientially hollow. This stems from a fundamental mismatch between what RLVR optimizes (\textbf{what to say}) and what users value (\textbf{how to say it}). While factual correctness is a necessary condition, it is not sufficient for high-quality audio interaction.

To bridge this gap, we introduce \textbf{Step-Audio-R1.5}, which complements RLVR with Reinforcement Learning from Human Feedback (RLHF). Rather than relying solely on binary correctness checks, we train a reward model on holistic human preference judgments over end-to-end interactions. This approach distills correctness, fluency, and emotional resonance into a unified supervisory signal, enabling the policy to escape the reward trap and optimize for overall response quality rather than isolated factual accuracy.

Comprehensive evaluations confirm that Step-Audio-R1.5 preserves the analytical reasoning cultivated by RLVR while substantially improving multi-turn interaction quality. On traditional reasoning benchmarks, the model remains highly competitive. We further evaluate its conversational capabilities on the AudioMultiChallenge~\cite{audiomc} benchmark, which rigorously tests four key dimensions of spoken dialogue — Inference Memory, Instruction Retention, Self Coherence, and Voice Editing — under naturalistic multi-turn conditions. In this demanding setting, Step-Audio-R1.5 demonstrates robust capabilities that rival or exceed those of leading commercial systems such as Gemini-2.5-Flash in key interaction dimensions. To our knowledge, Step-Audio-R1.5 is the first audio reasoning model to systematically integrate RLHF, demonstrating that the verifiable reward trap is not an inherent limitation of audio CoT, but an artifact of an impoverished reward signal that human feedback can effectively resolve.

\section{Architecture}
\label{sec:architecture}

Building upon the structural foundation established by Step-Audio-R1, Step-Audio-R1.5 employs a streamlined architecture explicitly tailored for extended audio-based reasoning. The model comprises three primary components: an audio encoder, an audio adaptor, and a Large Language Model (LLM) decoder.

The acoustic front-end utilizes the Qwen2 audio encoder~\cite{chu2024qwen2}, which is extensively pretrained on diverse speech and audio understanding tasks. Operating at a frame rate of 25 Hz, the encoder is kept strictly frozen throughout the training pipeline to preserve its robust auditory perception. To bridge the continuous acoustic modality with the discrete textual space, an audio adaptor applies a temporal downsampling rate of 2. This effectively compresses the latent representations to 12.5 Hz, mitigating sequence length explosion~\cite{dao2023flashattention2,zhang2025mamba,zhang2025rethinking,zhang2025auto} during complex, multi-turn interactions.

The core reasoning engine is an LLM decoder initialized from Qwen2.5 32B~\cite{team2024qwen2}. It directly ingests the downsampled audio features to generate purely textual outputs. To support sophisticated Chain-of-Thought (CoT) reasoning, the generation process is structurally partitioned: the decoder is prompted to first synthesize explicit intermediate reasoning traces before auto-regressively generating the final reply. This decoupling of internal analysis and external response is critical, as it forms the architectural basis for seamlessly integrating Reinforcement Learning from Human Feedback (RLHF).
\section{Training Method}
\label{sec:training_method}

\subsection{Audio-Centric Mid-Training}
Given the base audio-language model $\pi_{\theta_0}$, we perform an audio-centric mid-training stage to strengthen audio understanding, audio-grounded reasoning, and general deliberative capability before post-training alignment. The training objective combines audio-grounded reasoning data with auxiliary text-only reasoning data under a unified supervised objective:
\begin{equation}
\mathcal{L}_{\mathrm{mid}}
=
\mathbb{E}_{(x,q,r,y)\sim\mathcal{D}_{\mathrm{audio}}}
\left[
\log \pi_{\theta}(r,y \mid x,q)
\right]
+
\mathbb{E}_{(q,r,y)\sim\mathcal{D}_{\mathrm{text}}}
\left[
\log \pi_{\theta}(r,y \mid q)
\right],
\end{equation}
where $(x,q,r,y)$ denotes audio-grounded samples with input audio $x$, associated textual context $q$, reasoning trace $r$, and response $y$, while $(q,r,y)$ denotes text-only samples with context $q$, reasoning trace $r$, and response $y$. Audio-grounded supervision is drawn from diverse, high-quality audio-centric data, allowing the model to build broad perceptual coverage and robust reasoning capability over acoustically grounded contexts. Complementarily, auxiliary text-only supervision provides high-quality reasoning traces and long-form deliberative structure, facilitating the transfer of these reasoning patterns to audio-grounded understanding and inference.

\subsection{Cold-start Supervised Fine-tuning}
We perform a cold-start supervised fine-tuning (SFT) stage to initialize the model for interaction-oriented alignment. Although mid-training improves audio-domain knowledge, perceptual capability, and general reasoning ability, it does not directly optimize the model for high-quality multi-turn interaction. As a result, strong audio understanding alone is insufficient to ensure natural, coherent, and instruction-sensitive dialogue behavior.

Rather than further expanding domain knowledge, cold-start SFT provides a supervised initialization for interaction-oriented behavior prior to preference-based optimization. Concretely, this stage emphasizes four aspects of interaction behavior: \emph{(1) multi-turn dialogue continuity}, the ability to maintain context and user constraints across turns; \emph{(2) instruction following}, the ability to respond consistently under user-specified requirements on content, format, and style; \emph{(3) response naturalness}, the ability to produce coherent and conversationally appropriate responses; and \emph{(4) interaction awareness}, the ability to respond robustly to follow-up questions, clarification requests, interruptions, and user-side revisions.

To support these objectives, cold-start SFT is constructed from instruction-rich, multi-turn conversational data that encourages the model to organize responses in a user-oriented manner rather than as isolated task outputs. This stage provides a stronger conversational initialization for the subsequent RLHF stage, allowing preference optimization to focus on refining holistic interaction quality rather than correcting basic dialogue behavior.

\subsection{RLHF with Rubric-based Generated Reward Model}

Multi-turn spoken interaction exhibits substantially heterogeneous optimization targets. Some behaviors are governed by explicit and localized constraints, such as content requirements, formatting specifications, persona settings, and instruction retention across turns. Others are inherently preference-driven and only weakly specifiable, including conversational naturalness, coherence under follow-up interaction, appropriateness of tone, and overall dialogue fluency. These objectives differ not only in form, but also in how they should be evaluated: some admit relatively clear criteria, whereas others are better captured through comparative preference judgments over complete responses.

To accommodate this heterogeneity, we adopt a unified RLHF framework based on a generated reward model that jointly supports rubric-guided evaluation and ordinary preference comparison. For samples with explicit evaluation criteria, the reward model conditions on task-specific rubrics to assess whether the response satisfies the intended requirements. For samples without such criteria, the model instead performs standard pairwise preference judgment against a reference response. Formally, let $\mathcal{H}_{1:T}=\{h_t\}_{t=1}^{T}$ denote a multi-turn dialogue history up to turn $T$, where each $h_t$ represents the full interaction context at turn $t$. Given $\mathcal{H}_{1:T}$, a policy response $y$, a reference response $y^{\mathrm{ref}}$, and an optional rubric $c$, the generated reward model produces a relative quality judgment
\begin{equation}
g = \mathcal{R}(\mathcal{H}_{1:T}, y, y^{\mathrm{ref}}; c), \qquad c \in \mathcal{C} \cup \{\varnothing\},
\end{equation}
where $c=\varnothing$ corresponds to ordinary pairwise preference comparison, while $c \neq \varnothing$ denotes rubric-conditioned evaluation. The judgment $g$ is then mapped to a scalar reward
\begin{equation}
r = \phi(g),
\end{equation}
which is used for subsequent policy optimization. We optimize the policy by maximizing a PPO-style objective,
\begin{equation}
\resizebox{0.9\hsize}{!}{%
$\displaystyle
\begin{aligned}
\mathcal{L}_{\mathrm{RLHF}}(\theta)
&= 
\mathbb{E}_{t}\left[
\min\left(
\rho_t(\theta)\hat{A}_t,\;
\mathrm{clip}\!\left(\rho_t(\theta), 1-\epsilon, 1+\epsilon\right)\hat{A}_t
\right)
\right]
-\beta\, D_{\mathrm{KL}}\!\left(
\pi_{\theta}(\cdot \mid \mathcal{H}_{1:T}, c)
\,\|\,
\pi_{\mathrm{ref}}(\cdot \mid \mathcal{H}_{1:T}, c)
\right)
\end{aligned}
$%
}
\end{equation}
where
\begin{equation}
\rho_t(\theta)
=
\frac{\pi_{\theta}(y_t \mid \mathcal{H}_{1:T}, c)}
{\pi_{\theta_{\mathrm{old}}}(y_t \mid \mathcal{H}_{1:T}, c)},
\end{equation}
$\hat{A}_t$ is the advantage estimated from the generated reward, and $\pi_{\mathrm{ref}}$ denotes the reference policy used for regularization. These two forms of supervision are optimized jointly rather than in separate stages, since their optimization directions can differ substantially; empirically, decoupled training tends to induce non-trivial forgetting, where later optimization on one interaction regime degrades behaviors acquired in the other. Joint optimization therefore provides a more stable route to aligning both instruction-sensitive and preference-sensitive aspects of multi-turn dialogue within a single policy.

Within this unified RLHF framework, supervision is instantiated through a generated reward model based on relative comparison. Instead of assigning an absolute quality score to each response, the reward model compares the policy response against a reference response under the same multi-turn dialogue context and produces a preference judgment according to their comparative quality. This relative reward formulation is better suited to spoken dialogue alignment, where many important aspects of interaction quality are difficult to calibrate with a single absolute score. By representing reward as a fine-grained relative preference signal with multiple ordinal levels, the model can capture different degrees of response quality beyond binary distinction, yielding a more discriminative supervision signal for policy optimization.
\section{Evaluation}
\subsection{Benchmarks}
To comprehensively evaluate Step-Audio-R1.5's reasoning and perception capabilities, we employ a suite of speech-to-text (S2T) benchmarks. S2T evaluation isolates the model's ability to understand and reason over acoustic signals by requiring text-based responses, enabling direct comparison with state-of-the-art large language models.

\paragraph{AudioMultiChallenge (Audio MC).}
AudioMultiChallenge \cite{audiomc} is a multi-turn benchmark that evaluates spoken dialogue systems on natural human interaction patterns, including interruptions, hesitations, and mid-utterance repairs. It measures performance across four dimensions: Inference Memory, Instruction Retention, Self Coherence, and Voice Editing, providing a comprehensive assessment of a model's ability to handle long-context dialogue, follow instructions over multiple turns, and maintain consistency under real-world conversational noise.

\paragraph{Step-Caption.}
Step-Caption is a newly proposed benchmark designed to evaluate the model's fine-grained audio description capability. The test set consists of 905 carefully curated audio samples sourced from YouTube and Bilibili, covering both single-speaker and multi-speaker scenarios primarily in Chinese and English. Each sample is annotated by human experts across 16 dimensions, including gender, age, speaking rate, rhythm, pitch, timbre, emotion, accent, and other paralinguistic features. The model is required to generate a natural language paragraph that comprehensively describes the speaker's vocal characteristics, with the prompt explicitly requesting analysis of all 16 dimensions. This benchmark specifically measures the model's ability to perceive and articulate acoustic attributes such as timbre, age, gender, and emotional state from raw audio.

\paragraph{Step-Dialogue-Understanding (Step-DU).}
While Step-Caption focuses on a comprehensive acoustic description, Step-Dialogue-Understanding evaluates the model's ability to answer specific questions about paralinguistic features in a conversational context. The test set consists of 87 samples recorded by diverse speakers, each directly asking about their own vocal characteristics, such as age, gender, speaking rate, or rhythm. The model must infer the correct answer solely from the acoustic signal, testing its perception and reasoning of paralinguistic cues in an interactive dialogue setting.

\paragraph{StepEval-Audio-Paralinguistic (Step-SPQA).}
StepEval-Audio-Paralinguistic was originally introduced as an AQAA (Audio Query–Audio Answer) benchmark in Step-Audio 2 \cite{wu2025step}. To ensure consistent text-based evaluation across all models in this work, we have converted it to the AQTA (Audio Query–Text Answer) format, while preserving the original audio understanding tasks.

\paragraph{Additional Public Benchmarks.}
In addition to the proposed benchmarks, we also report results on widely adopted public benchmarks to enable broad comparison with existing models. These include MMSU \cite{wang2025mmsu} and MMAU \cite{sakshi2024mmau} for expert-level audio understanding and reasoning, Big Bench Audio\footnote{\url{https://huggingface.co/datasets/ArtificialAnalysis/big_bench_audio}} for complex multi-step logical reasoning from audio, and Spoken MQA \cite{wei2025towards} for mathematical reasoning with verbally expressed problems.

\subsection{Experimental Results}

To ensure a fair and consistent comparison, we evaluated all baseline models using their official APIs through our own unified evaluation framework, rather than relying on previously reported numbers. This approach guarantees that all results are directly comparable under identical conditions. The baseline models include the Gemini family (Gemini 3 Flash and Gemini 3 Pro)\footnote{\url{https://blog.google/technology/developers/gemini-3-pro-vision/}} and the Qwen family (qwen3.5-omni-flash and qwen3.5-omni-plus)~\cite{team2026qwen3}.

\begin{table}[ht]
\caption{Performance comparison on speech-to-text benchmarks. Avg. is calculated over all benchmarks for each model. Best results in bold, second-best underlined.}
\label{tab:s2t}
\renewcommand\arraystretch{1.1}
\begin{center}
\resizebox{\textwidth}{!}{%
\begin{tabular}{lccccccccc}
\toprule
Model & \textbf{Avg.} & \textbf{Audio MC} & \textbf{Big Bench} & \textbf{MMSU} & \textbf{MMAU} & \textbf{Spoken MQA} & \textbf{Step-Caption} & \textbf{Step-DU} & \textbf{Step-SPQA} \\
\midrule
Gemini 3 Flash & 77.56 & \underline{56.42} & 96.80 & 76.64 & 75.90 & 95.37 & 65.12 & 80.46 & 73.80 \\
Gemini 3 Pro & \textbf{79.67} & \textbf{66.37} & \textbf{99.40} & \textbf{83.70} & \textbf{79.80} & \textbf{96.56} & \textbf{75.55} & 72.41 & 63.60 \\
qwen3.5-omni-flash & 70.55 & 25.44 & 59.59 & 72.50 & 77.20 & 93.39 & 73.57 & \underline{83.91} & 78.80 \\
qwen3.5-omni-plus & 75.77 & 39.38 & 73.03 & \underline{82.74} & \underline{79.60} & \underline{96.03} & \underline{74.93} & \textbf{85.63} & \underline{74.80} \\
\midrule
Step-Audio-R1 & 72.50 & 24.61 & 98.29 & 75.68 & 77.00 & 95.06 & 70.60 & 64.37 & 74.36 \\
\midrule
\textbf{Step-Audio-R1.5} & \underline{77.97} & 41.15 & \underline{98.30} & 79.03 & 77.90 & 93.74 & 71.48 & 82.76 & \textbf{79.40} \\
\bottomrule
\end{tabular}
}%
\end{center}
\end{table}

As shown in Table~\ref{tab:s2t}, Step-Audio-R1.5 achieves an average score of 77.97, ranking second among all evaluated models and demonstrating competitive performance against much larger proprietary models. Notably, despite having only 32B parameters, Step-Audio-R1.5 attains 41.15 on Audio MC, a highly competitive result that trails only the Gemini family models. Across all benchmarks, Step-Audio-R1.5 maintains balanced performance, achieving a significant average score improvement of 5.47 points over its predecessor Step-Audio-R1 (72.50). This gain is primarily driven by substantial advances on complex tasks requiring multi-turn and long-context understanding, as evidenced by the Audio MC benchmark, while its performance on perceptual benchmarks also shows broad improvements: substantial gains on Step-DU (+18.39) and Step-SPQA (+5.04), alongside a modest gain on Step-Caption (+0.88). These results collectively validate the effectiveness of our architecture and training pipeline.
\section{Conclusion}
\label{sec:conclusion}

The mechanical, emotionally flat responses observed in early audio reasoning models are not an inherent limitation of the Chain-of-Thought process, but rather an artifact of the verifiable reward trap. In this work, we demonstrate that heavily optimizing for isolated semantic correctness via RLVR structurally blinds models to the multidimensional nuances of genuine human interaction. Step-Audio-R1.5 breaks this trade-off by systematically integrating Reinforcement Learning from Human Feedback (RLHF), leveraging a decoupled generation architecture and a rubric-guided preference reward model. By realigning the optimization objective from merely \emph{what to say} to holistically \emph{how to say it}, Step-Audio-R1.5 substantially improves multi-turn conversational quality while preserving analytical rigor. This work provides a critical insight for the evolution of audio language models: as acoustic understanding matures, the next frontier of artificial audio intelligence lies not in reducing continuous sensory inputs to discrete factual puzzles, but in aligning model behavior with the rich, empathetic dynamics of natural spoken dialogue.

\section{Contributors}

\vspace{0.5em}
\textbf{Core Contributors:}
\textbf{Yuxin Zhang}$^{1,4}$,
\textbf{Xiangyu Tony Zhang}$^{3}$,
\textbf{Daijiao Liu}$^{1,3}$,
\textbf{Fei Tian}$^{1,*,\dagger}$,
\textbf{Yayue Deng}$^{1}$,
\textbf{Jun Chen}$^{1}$,
\textbf{Qingjian Lin}$^{1}$

\vspace{0.5em}
\textbf{Contributors:}
\textbf{Haoyang Zhang}$^{1,2}$,
\textbf{Yuxin Li}$^{1,2}$,
\textbf{Jinglan Gong}$^{1}$, 
\textbf{Yechang Huang}$^{1}$, 
\textbf{Liang Zhao}$^{1}$, 
\textbf{Chengyuan Yao}$^{1}$, 
\textbf{Hexin Liu}$^{2}$, 
\textbf{Eng Siong Chng}$^{2}$,
\textbf{Xuerui Yang}$^{1}$,
\textbf{Gang Yu}$^{1}$,
\textbf{Xiangyu Zhang}$^{1}$,
\textbf{Daxin Jiang}$^{1}$

\vspace{0.5em}
$^1$StepFun \hspace{2em} $^2$Nanyang Technological University \hspace{2em} $^3$University of New South Wales \hspace{2em} $^4$Shanghai Jiao Tong University \\
$^*$Corresponding authors: \texttt{tianfei@stepfun.com} \hspace{3em} $^\dagger$Project Leader



\setlength{\bibsep}{0.5\baselineskip}
\bibliography{references}

\end{document}